\documentclass[fleqn,usenatbib]{mnras}

\usepackage{newtxtext,newtxmath}

\usepackage[T1]{fontenc}

\DeclareRobustCommand{\VAN}[3]{#2}
\let\VANthebibliography\thebibliography
\def\thebibliography{\DeclareRobustCommand{\VAN}[3]{##3}\VANthebibliography}

\usepackage{graphicx}	
\usepackage{amsmath}	

\title[Energetics of the gamma-ray emission from pulsars]{Energetic requirements of the gamma-ray emission from pulsars: A nonparametric analysis of the data in the Fermi-LAT 12-Year Catalog}

\author[H. Ardavan]{
Houshang Ardavan\thanks{E-mail: ardavan@ast.cam.ac.uk}
\\
Institute of Astronomy, University of Cambridge, Madingley Road, Cambridge CB3 0HA, UK\\
}

\date{Accepted XXX. Received YYY; in original form ZZZ}

\pubyear{2022}

\begin{document}
\label{firstpage}
\pagerange{\pageref{firstpage}--\pageref{lastpage}}
\maketitle

\begin{abstract}
The prevalent view that the radio-loud gamma-ray pulsars have gamma-ray luminosities that exceed their radio luminosities by several orders of magnitude is based on the assumption that the decay with distance of their gamma-ray fluxes obeys the inverse-square law as does that of their radio fluxes.  The results presented here, of testing the hypothesis of independence of luminosities and distances of gamma-ray pulsars by means of the Efron--Petrosian statistic, do not uphold this assumption however: they imply that the observational data in the Fermi-LAT 12-Year Catalog are consistent with the dependence $S\propto D^{-3/2}$ of the flux densities $S$ of the gamma-ray pulsars on their distances $D$ at substantially higher levels of significance than they are with the dependence $S\propto D^{-2}$.  These results, which were theoretically predicted in Ardavan (2021, MNRAS, 507, 4530), are not incompatible with the requirements of the conservation of energy because the radiation process by which the superluminally moving current sheet in the magnetosphere of a neutron star has been shown to generate the slowly decaying gamma-ray pulses is intrinsically transient: the difference in the fluxes of power across any two spheres centred on the star is balanced by the change with time of the energy contained inside the shell bounded by those spheres.  Once the over-estimation of their values is rectified, the luminosities of gamma-ray pulsars turn out to have the same range of values as do the luminosities of radio pulsars.  This conclusion agrees with that reached earlier on the basis of the smaller data set in the Second Fermi-LAT Catalog of Gamma-ray Pulsars.
\end{abstract}




\section{Introduction}
\label{sec:introduction}

Numerical computations based on the force-free and particle-in-cell formalisms have now firmly established that the magnetosphere of a non-aligned neutron star entails a current sheet outside its light cylinder whose rotating distribution pattern moves with linear speeds exceeding the speed of light in vacuum~\citep[see the review article][and the references therein]{Philippov2022}.  A study of the characteristics of the radiation that is generated by this superluminally moving current sheet has in turn provided an all-encompassing explanation for the salient features of the radiation received from pulsars: its brightness temperature, polarization, spectrum and profile with microstructure and with a phase lag between the radio and gamma-ray peaks~\citep{Ardavan2021, Ardavan2022a}.  A fit to the exceptionally broad gamma-ray spectrum of the Crab pulsar, for example, is provided by the spectral distribution function of this radiation for the first time~\citep{Ardavan2022d}. 

The radiation field generated by a constituent volume element of the current sheet in the magnetosphere of a neutron star embraces a synergy between the superluminal version of the field of synchrotron radiation and the vacuum version of the field of \v{C}erenkov radiation.  Once superposed to yield the emission from the entire volume of the source, the contributions from the volume elements of the current sheet that approach the observation point with the speed of light and zero acceleration at the retarded time interfere constructively and form caustics in certain latitudinal directions relative to the spin axis of the neutron star.  The waves that embody these caustics are more focused the further they are from their source: as their distance from their source increases, two nearby stationary points of their phases draw closer and eventually coalesce at infinity.  As a result, flux densities of the pulses that are generated by this current sheet diminish with the distance $D$ from the star as $D^{-3/2}$ (rather than $D^{-2}$) in the latitudinal directions along which the pulses are most tightly focused~\citep[][Section 5.5]{Ardavan2021}.  By virtue of their extremely narrow peaks in the time domain, such pulses in addition have broad spectra that encompass X-ray and gamma-ray frequencies~\citep[][Table~1 and Section~5.4]{Ardavan2021}.  

It is expected, therefore, that the class of progenitors of the non-spherically decaying radiation in question would include, {\it inter alia}, the gamma-ray pulsars. 

To see whether this expectation is supported by the observational data on fluxes and distances of gamma-ray pulsars, we analyse the latest version of these data~\citep{Abdollahi2022} here on the basis of the fact that, in a statistical context, luminosities and distances of the sources of a given type of radiation represent two {\it independent} random variables.  A generalization of the nonparametric rank methods for testing the independence of two random variables~\citep{Hajek} to cases of truncated data, such as the flux-limited data on gamma-ray pulsars, is the method developed by~\citet{EF1992,EF1994}: a method that has been widely used in astrophysical contexts~\citep[see e.g.][and the references therein]{Bryant}.   

From the raw data on fluxes $S$ of gamma-ray pulsars in the Fermi-LAT 12-Year Catalog\footnote{{\url{https://fermi.gsfc.nasa.gov/ssc/data/access/lat/12yr_catalog/}}}, the distances $D$ of those of them that are listed in the ATNF Pulsar Catalogue\footnote{{\url{http://www.atnf.csiro.au/research/pulsar/psrcat}}}, and the candidate decay rates of flux density with distance ($S\propto D^{-\alpha}$ for various values of $\alpha$) we compile a collection of data sets on the prospective luminosities of these pulsars (Section~\ref{sec:data}).  We then test the hypothesis of independence of luminosity and distance by evaluating the Efron--Petrosian statistic (Section~\ref{subsec:statistic}) for the data sets on prospective luminosities and pulsar distances choosing a wide range of values for the flux threshold (i.e. the truncation boundary below which the data set on $S$ may be incomplete).  The resulting values of the Efron--Petrosian statistic for differing values of $\alpha$ in each case determine the significance levels at which the hypothesis of independence of luminosity and distance (and hence a given value of $\alpha$) can be rejected (Sections~\ref{subsec:millisecond} and~\ref{subsec:young}).  

We analyse the two homogeneous data sets on the young and the millisecond pulsars in the Fermi-LAT 12-Year Catalog separately, each with both the NE2001~\citep{NE2001} and the YMW16~\citep{YMW16} distances.  We also assess the effects of random and (if any) systematic errors in the estimates of flux on the test results by means of a Monte Carlo simulation (Section~\ref{subsec:errors}). Whether the sizes of the currently available data sets are sufficient for the purposes of the present rank analysis is ascertained in Section~\ref{subsec:size} where we augment these data sets by the inclusion of their permissible permutations.

\begin{figure*}
\centerline{\includegraphics[width=17cm]{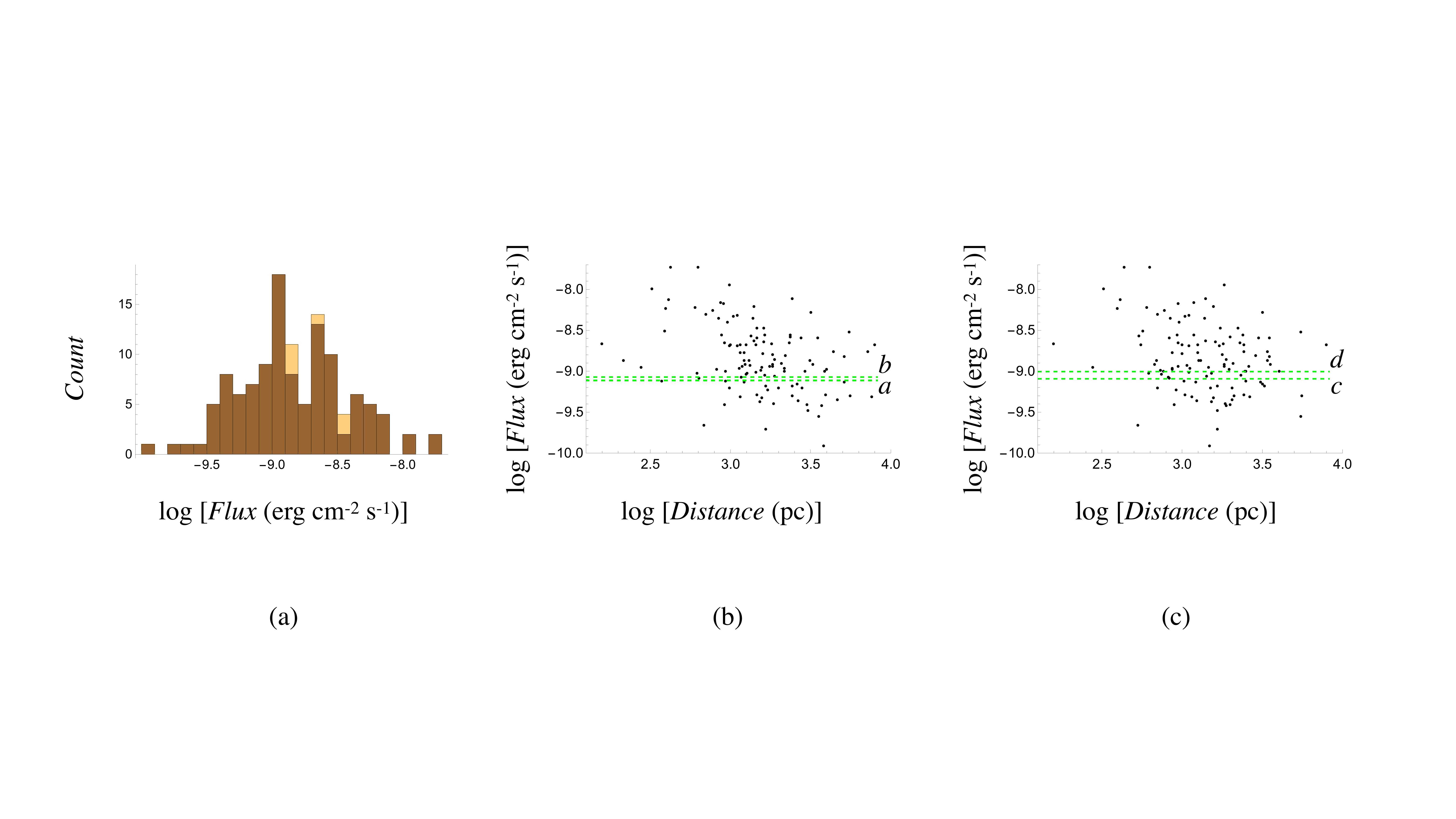}}
\caption{(a)  Histogram of the $120$ millisecond gamma-ray pulsars listed in the Fermi-LAT 12-Year Catalog.   The bins in darker brown contain the $114$ of these pulsars whose distances are listed in the ATNF Pulsar Catalogue.  (b) Distribution of logarithm of flux versus logarithm of distance for the millisecond pulsars in the darker brown bins based on the YMW16 distances in the ATNF Catalogue.  The dashed lines $a$ and $b$ respectively designate the flux thresholds $\log S_{\rm th}=-9.112$ and $\log S_{\rm th}= -9.07$, i.e.\ the thresholds marked in Fig.~\ref{NLF4}a.  (c) Distribution of logarithm of flux versus logarithm of distance for the same set of pulsars based on the NE2001 distances in the ATNF Catalogue.  The dashed lines $c$ and $d$ respectively designate the flux thresholds $\log S_{\rm th}=-9.092$ and $\log S_{\rm th}= -9.005$, i.e. the thresholds marked in Fig.~\ref{NLF4}b.}
\label{NLF1}
\end{figure*}

\begin{figure*}
\centerline{\includegraphics[width=18cm]{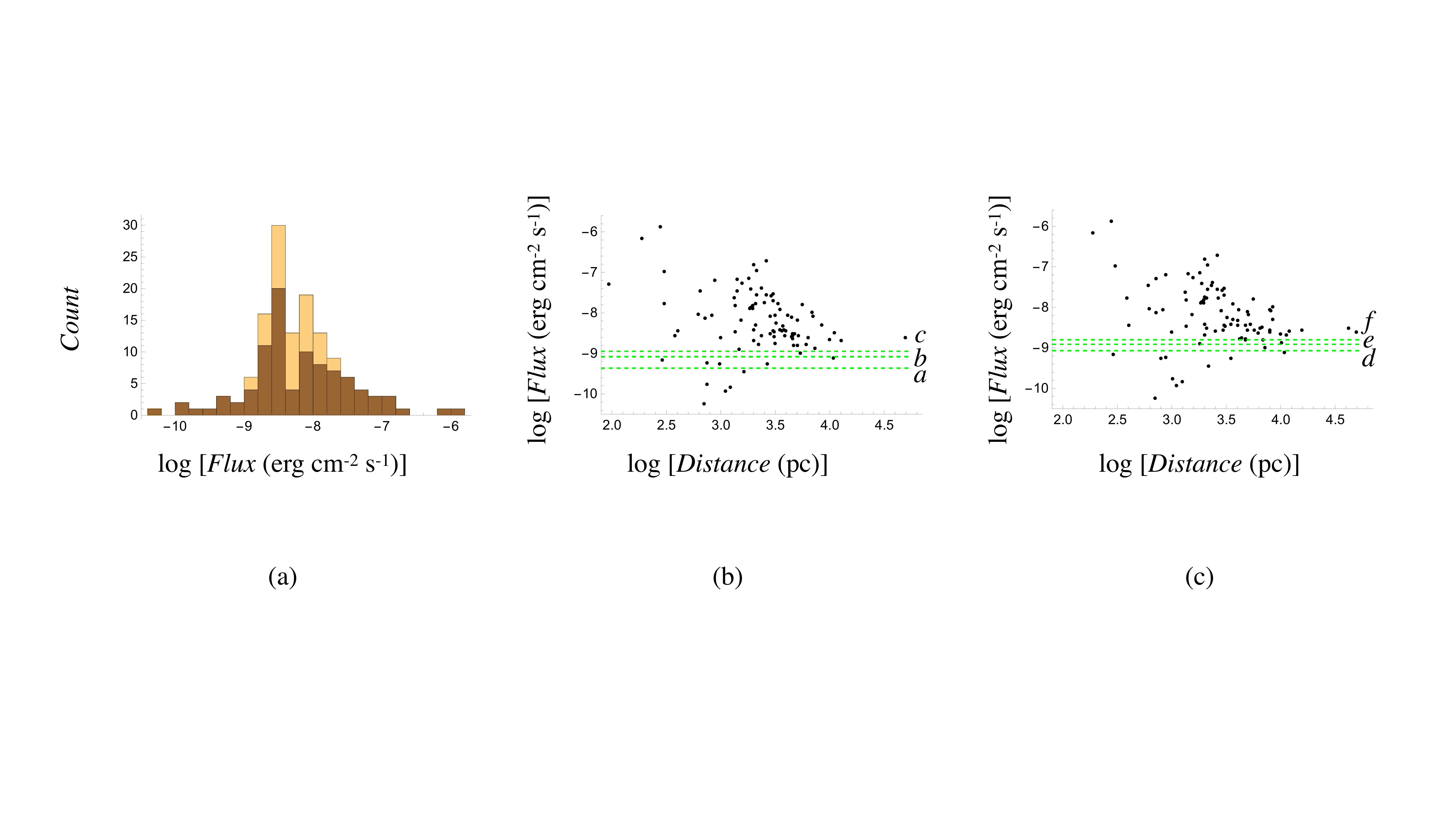}}
\caption{(a)  Histogram of the $135$ young gamma-ray pulsars listed in the Fermi-LAT 12-Year Catalog.   The bins in darker brown contain the $93$ of these pulsars whose distances are listed in the ATNF Pulsar Catalogue.  (b) Distribution of logarithm of flux versus logarithm of distance for the young pulsars in the darker brown bins based on the YMW16 distances in the ATNF Catalogue.  The dashed lines $a$, $b$ and $c$ respectively designate the flux thresholds $\log S_{\rm th}=-9.365$,  $\log S_{\rm th}=-9.08$ and $\log S_{\rm th}= -8.95$, i.e.\ the thresholds marked in Fig.~\ref{NLF6}a.  (c) Distribution of logarithm of flux versus logarithm of distance for the same set of pulsars based on the NE2001 distances in the ATNF Catalogue.  The dashed lines $d$, $e$ and $f$ respectively designate the flux thresholds $\log S_{\rm th}=-9.066$, $\log S_{\rm th}=-8.911$ and $\log S_{\rm th}= -8.8$, i.e.\ the thresholds marked in Fig.~\ref{NLF6}b.}
\label{NLF2}
\end{figure*}

\section{Observational data}
\label{sec:data}

The Fermi-LAT 12-Year Catalog lists the observational data on $120$ millisecond and $135$ young gamma-ray pulsars.  Of these, there are $6$ millisecond and $42$ young pulsars with no distance estimates in the ATNF Pulsar Catalogue.  Histograms of the fluxes of these two sets of pulsars are shown in Figs.~\ref{NLF1}a and \ref{NLF2}a.  The histogram bins in darker brown contain the pulsars for which both fluxes and distances are known.  Figures~\ref{NLF1} and \ref{NLF2} also show the distributions of logarithm of flux (in units of erg cm$^{-2}$ s$^{-1}$) versus logarithm of distance (in units of pc) for these two sets of pulsars.  Parts (b) and (c) of these figures are based on the distances in the ATNF Catalogue that are listed under YMW16 and NE2001, respectively.  The dashed lines in them each designate the value of a flux threshold $S_{\rm th}$ below which the plotted data set may be incomplete.

\begin{figure*}
\centerline{\includegraphics[width=16cm]{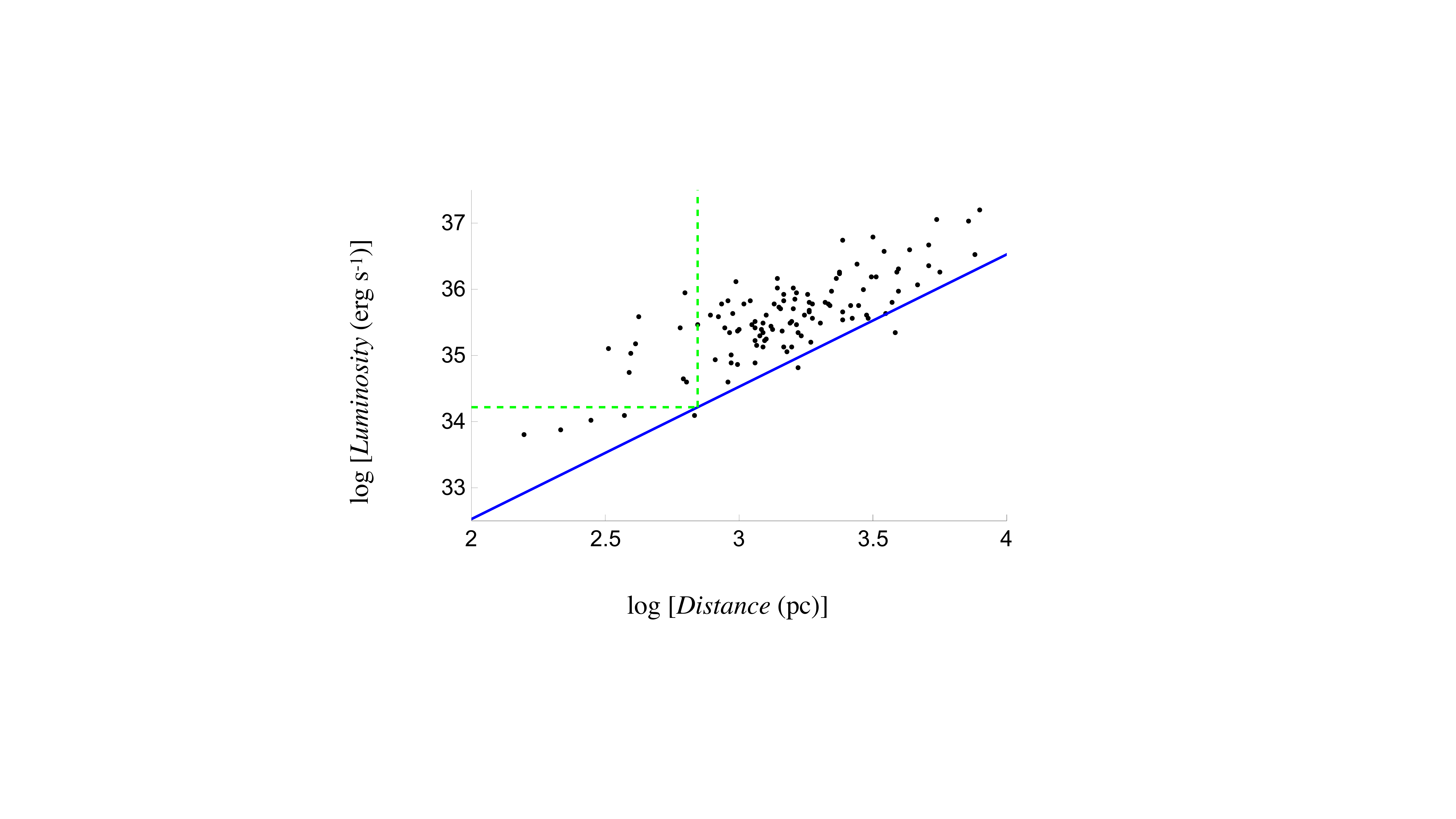}}
\caption{The luminosity-distance data set that follows from the flux-distance data set of Fig.~\ref{NLF1}b and equation (\ref{E1}) for $\alpha=2$.  The solid line (in blue) is the image $L=L_{\rm th}(D)$ of the flux threshold $\log S_{\rm th}=-9.55$.  Those elements of this data set that lie within (and on the boundary of) the rectangular area bounded by the vertical axis and the vertical and horizontal dashed lines (in green) comprise the set comparable to the element $(2.84, 35.46)$ on the vertical dashed line.} 
\label{NLF3}
\end{figure*}

The isotropic gamma-ray luminosity of each pulsar is given, in terms of its gamma-ray flux density $S$ and its distance $D$, by 
\begin{equation}
L=4\pi \ell^2(D/\ell)^\alpha S,
\label{E1}
\end{equation}
where $\alpha=2$ if $S$ diminishes with distance as predicted by the inverse-square law and $\ell$ is a constant with the dimension of length whose value only affects the scale of $L$.  The corresponding distribution of the logarithm of $L$ (in units of erg s$^{-1}$) versus logarithm of $D$ (in units of pc) for the data set shown in Fig.~\ref{NLF1}b is plotted in Fig.~\ref{NLF3} for $\alpha=2$.  The solid line in Fig.~\ref{NLF3} shows an example of a flux threshold: that corresponding to the threshold $S=2.82\times10^{-10}$ erg cm$^{-2}$ s$^{-1}$ which excludes three elements of this data set.

\section{Testing the hypothesis that the data sets on luminosity and distance are independent}
\label{sec:test}
\subsection{The Efron--Petrosian rank statistic}
\label{subsec:statistic}

In this section we only outline the procedure by which the Efron--Petrosian statistic is calculated for two given data sets; lucid expositions of the theoretical basis of this procedure can be found in~\citet{EF1992},~\citet{Maloney1999} and~\citet{Petrosian2002}. 

If we let $S_{\rm th}$ stand for the threshold value of flux density, then the corresponding truncation boundary for the values of luminosity (e.g. that shown as a solid line in Fig.~\ref{NLF3}) is given, according to equation~(\ref{E1}), by $L=L_{\rm th}(D)$ with
\begin{equation}
\log L_{\rm th}=\log[4\pi(3.085\times10^{18})^2\ell^{2-\alpha} S_{\rm th}]+\alpha\log D,
\label{E2}
\end{equation}
in which the numerical factor $3.085\times10^{18}$ converts the units of $D$ and $\ell$ from pc to cm.  The data set on luminosity is thus regarded as complete only in the sector $\log L\ge \log L_{\rm th}$ of the $(\log D,\log L)$ plane.

The set {\it comparable} to any given element $(\log D_i,\log L_i)$ of the bivariate distance-luminosity data set (such as the data set plotted in Fig.~\ref{NLF3}) is defined to comprise all those elements for which
\begin{equation}
\log D\le \log D_i,\qquad i=1,\cdots n,
\label{E3}
\end{equation}  
and
\begin{equation}
\log L\ge\log[4\pi(3.085\times10^{18})^2\ell^{2-\alpha} S_{\rm th}]+\alpha\log D_i,
\label{E4}
\end{equation}
where $n$ is the number of elements in the part of the data set that is not excluded by the chosen flux threshold.  For instance, the set comparable to the data point $(2.84, 35.46)$ on the green dashed line in Fig.~\ref{NLF3} consists of the elements of the data set that lie within (and on the boundary of) the rectangular region delineated by the vertical axis and the dashed lines coloured green in this figure.  We denote the number of elements in the set comparable to $(\log D_i,\log L_i)$ by $N_i$.

To determine the {\it rank} ($1\le R_i\le N_i$) of the element $(\log D_i,\log L_i)$, we now order the $N_i$ elements of its comparable set by the ascending values of their coordinates $\log L_i$ and equate $R_i$ to the position at which $(\log D_i,\log L_i)$ appears in the resulting ordered list.  Coordinates of the elements of the bivariate data set $(\log D,\log L)$ are in the present case all distinct.  So, a rank $R_i$ can be assigned to every element of this set unambiguously.  

The Efron--Petrosian {\it statistic} is given by
\begin{equation}
\tau=\frac{\sum_{i=1}^n(R_i-E_i)}{\sqrt{\sum_{i=1}^nV_i}},
\label{E5}
\end{equation}
in which
\begin{equation}
E_i=\textstyle{\frac{1}{2}}(N_i+1),\qquad V_i=\textstyle\frac{1}{12}(N_i^2-1).
\label{E6}
\end{equation}
Note that the value of $\tau$ is independent of that of the scale factor $\ell$ that appears in equation~(\ref{E1}): the value of the Efron--Petrosian statistic does not change if we make monotonically increasing transformations on the values of distance and/or flux~\citep{EF1992}.  

The hypothesis of independence is rejected when the value of $\tau$ falls in one of the tails of a Gaussian distribution whose mean is $0$ and whose variance is $1$.  More precisely, the hypothesis of independence is rejected when the value of
\begin{equation}
p=(2/\pi)^{1/2}\int_{\vert\tau\vert}^\infty\exp(-x^2/2){\rm d}x={\rm erfc}(\vert\tau\vert/\sqrt{2})
\label{E7}
\end{equation}
is smaller than an adopted significance level between $0$ and $1$, where ${\rm erfc}$ denotes the complementary error function.  If $\tau$ equals zero, for instance, then $p$ would assume the value one and the hypothesis in question cannot be rejected at any significance level.

\subsection{Test results for the millisecond pulsars}
\label{subsec:millisecond}

\begin{figure*}
\centerline{\includegraphics[width=17cm]{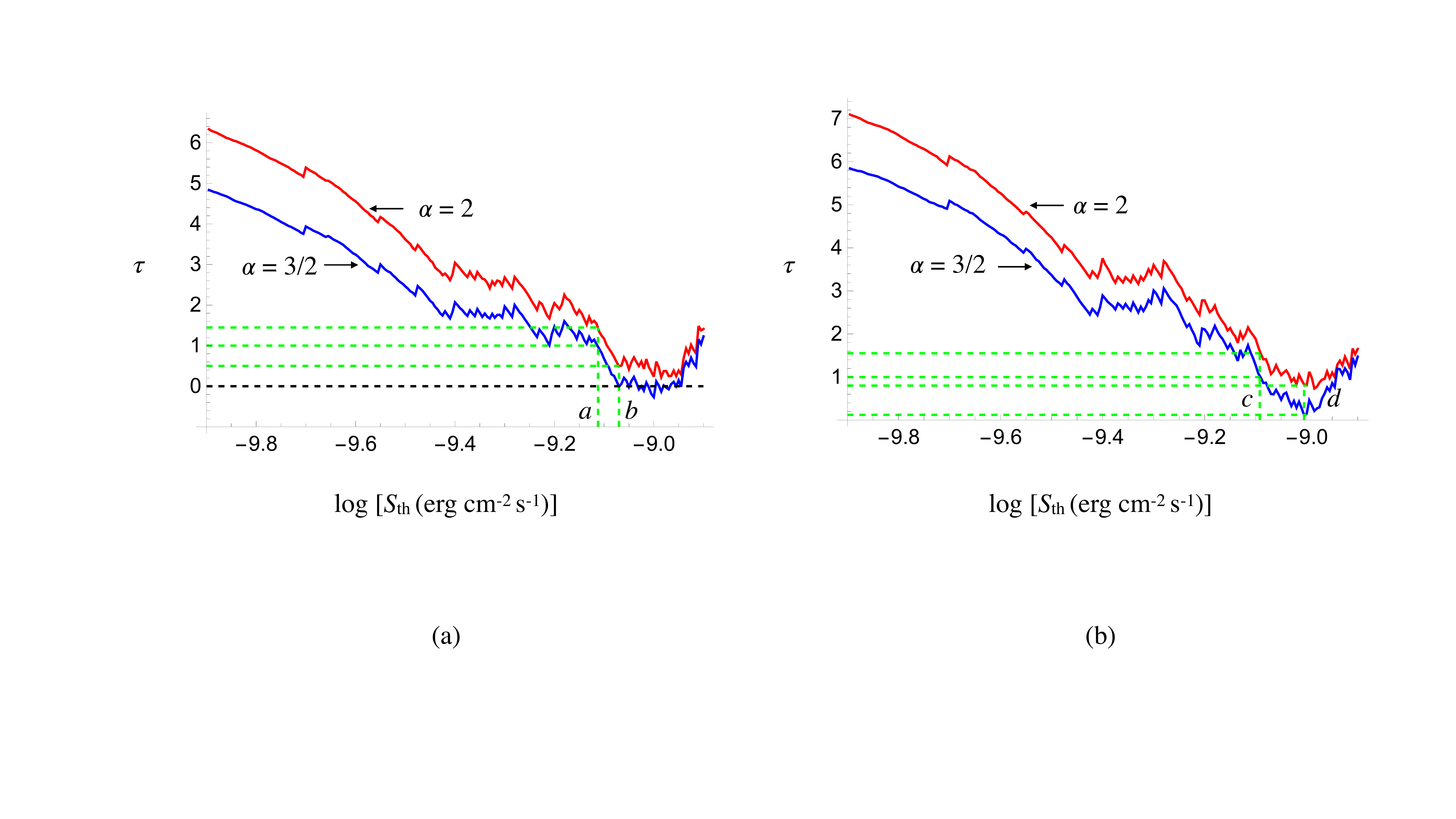}}
\caption{The Efron--Petrosian statistic $\tau$ versus the logarithm of the flux threshold $S_{\rm th}$ for the millisecond pulsars.  The two curves are plotted for the values $2$ (coloured red) and $3/2$ (coloured blue) of $\alpha$.  (a) For the distances listed under YMW16 in the ANTF Catalogue.  The dashed vertical lines $a$ and $b$ (in green) designate the same thresholds as those in Fig.~\ref{NLF1}b, i.e.\ $\log S_{\rm th}=-9.112$ and $\log S_{\rm th}=-9.07$, respectively.  The values of $\tau$ corresponding to threshold $a$ are $1$ for $\alpha=3/2$ and $1.45$ for $\alpha=2$.  The values of $\tau$ corresponding to threshold $b$ are $0$ for $\alpha=3/2$ and $0.5$ for $\alpha=2$.  (b)  For the distances listed under NE2001 in the ANTF Catalogue.  The dashed vertical lines $c$ and $d$ (in green) designate the same thresholds as those in Fig.~\ref{NLF1}c, i.e.\ $\log S_{\rm th}=-9.092$ and $\log S_{\rm th}=-9.005$, respectively.  The values of $\tau$ corresponding to threshold $c$ are $1$ for $\alpha=3/2$ and $1.55$ for $\alpha=2$.  The values of $\tau$ corresponding to threshold $d$ are $0.114$ for $\alpha=3/2$ and $0.822$ for $\alpha=2$.}
\label{NLF4}
\end{figure*}

\begin{figure*}
\centerline{\includegraphics[width=17cm]{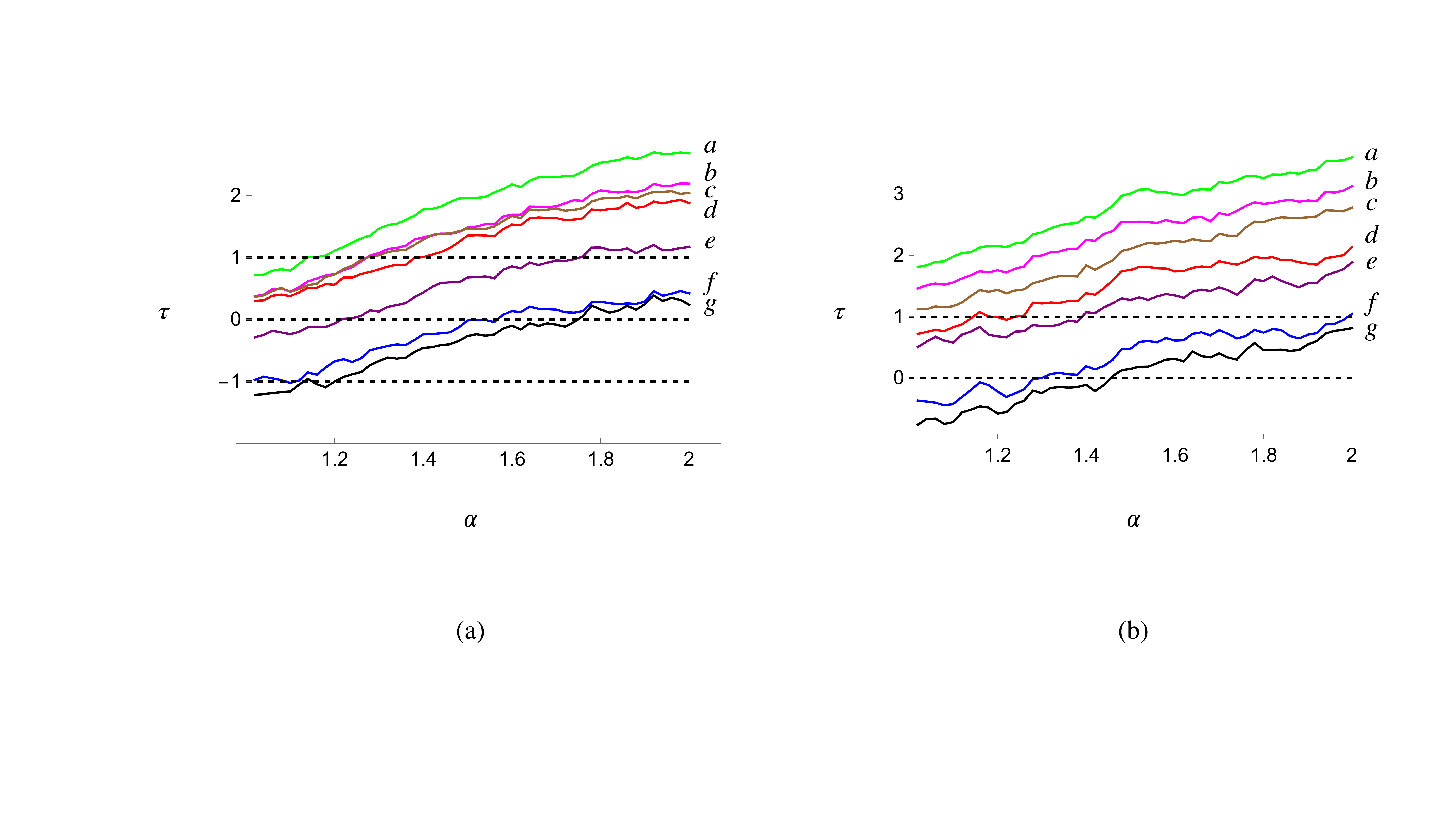}}
\caption{Contours of the function $\log S_{\rm th}(\alpha, \tau)$ over $-9.3\le\log S_{\rm th}\le-9.0$ at intervals of $0.05$ for the millisecond pulsars.  (a)  Based on the distances listed under YMW16 in the ANTF Catalogue and for the following values of $\log S_{\rm th}$: -9.3 ($a$, green), -9.25 ($b$, magenta), -9.2 ($c$, brown) , -9.15 ($d$, red), -9.1 ($e$, purple), -9.05 ($f$, blue), -9 ($g$, black). (b) The same as in part (a) but based on the distances listed under NE2001 in the ATNF Catalogue.}
\label{NLF5}
\end{figure*}

We have used the data on millisecond pulsars (Fig.~\ref{NLF1}) to evaluate the expression in equation~(\ref{E5}) for the Efron--Petrosian statistic $\tau$ as a function of the flux threshold $S_{\rm th}$, once letting $\alpha=2$ and another time letting $\alpha=3/2$.  The results that follow from the two alternative data sets depicted in Figs~\ref{NLF1}b and~\ref{NLF1}c (i.e.\ from YMW16 and NE2001 distances) are plotted in parts (a) and (b) of Fig.~\ref{NLF4}, respectively.  The flux thresholds designated by the letters $a$ to $d$ in these figures are the ones for which the value of $\tau$ for $\alpha=3/2$ is $1$ (thresholds $a$ and $c$), $0$ (threshold $b$) or minimum (threshold $d$).  The corresponding values of $\tau$ for $\alpha=2$ at the same flux thresholds are listed, together with their associated $p$ values, in Table~\ref{T1}.  Dependence of the Efron--Petrosian statistic $\tau$ on the exponent $\alpha$ at fixed values of the flux threshold $S_{\rm th}$ is shown in Fig.~\ref{NLF5}. 

It can be seen from Figs~\ref{NLF4} and \ref{NLF5} that the values of $\vert\tau\vert$ for $\alpha=2$ lie above those for $\alpha=3/2$ throughout the displayed thresholds.  In the case of threshold $a$ in Fig.~\ref{NLF4}a, which excludes $30$ of the $114$ elements in the data set on millisecond pulsars, for example, the $p$-values (defined in equation~\ref{E7}) for $\alpha=3/2$ and $\alpha=2$ are respectively given by $0.317$ and $0.147$: while the hypothesis of independence of luminosity and distance can be rejected at any significance level exceeding $14.7\%$ when $\alpha=2$, the rejection of this hypothesis requires a significance level exceeding $31.7\%$ when $\alpha=3/2$.  For threshold $b$ in Fig.~\ref{NLF4}a, which excludes $31$ of the $114$ elements in the data set, the $p$-values corresponding to $\alpha=3/2$ and $\alpha=2$ are respectively given by $1$ and $0.617$: while it can be rejected at any significance level exceeding $61.7\%$ when $\alpha=2$, the present hypothesis cannot be rejected at even a $100\%$ significance level when $\alpha=3/2$.  Similar results are obtained for the flux thresholds $c$ and $d$ in Fig.~\ref{NLF4}b (see Table~\ref{T1}).

There is an upper limit to how high the chosen value of the flux threshold $S_{\rm th}$ can be.  The Kolmogorov--Smirnov statistic $p_{KS}$ yields the probability that the truncated data set with the elements $S\ge S_{\rm th}$ and the uncut $114$-element data set depicted in Fig.~\ref{NLF1} are drawn from the same distribution: from the unknown distribution that is complete over all values of the flux density.  Evaluation of $p_{KS}$ for the millisecond pulsars shows that the value of the Kolmogorov--Smirnov statistic for any truncated data set with a flux threshold larger than $\log S_{\rm th}=-8.9$ is lower than $3.54\times10^{-9}$.  In other words, any truncated data set for which the value of $S_{\rm th}$ lies outside the range of flux thresholds plotted in Fig.~\ref{NLF4} is extremely unlikely to have the same origin as the uncut $114$-element data set provided by the observation of millisecond pulsars.

\begin{table*}
\centering
\caption{The Efron--Petrosian statistic $\tau$ and its associated $p$ and $p_{\rm perm}$ values for $\alpha=1.5$ and $2$ versus the flux threshold $S_{\rm th}$ for the millisecond pulsars.  The rows YMW16 and NE2001 refer to the alternative distances listed in the ATNF Pulsar Catalogue and the letters $a$--$d$ refer to the flux thresholds shown in Fig.~\ref{NLF4}.}
\label{tab:landscape}
\begin{tabular}{lccccccc}
\hline
${}$ & $\log [S_{\rm th}$ & $\tau\vert_{\alpha=1.5}$  & $p\vert_{\alpha=1.5}$ & $p_{\rm perm}\vert_{\alpha=1.5}$ & $\tau\vert_{\alpha=2}$ & $p\vert_{\alpha=2} $ & $p_{\rm perm}\vert_{\alpha=2} $\\
 {} & (erg cm$^{-2}$ s$^{-1})]$ & & & & & &\\
\hline
 {YMW16} & -9.112 ($a$)  & 1 & 0.317 & 0.311 & 1.45 & 0.147 & 0.150\\
 {} & -9.070 ($b$) &  0 & 1 & 1 &  0.5     & 0.617 & 0.605\\
 \hline
 {NE2001} & -9.092 ($c$) & 1 & 0.317 & 0.258  & 1.55 & 0.121 & 0.116\\ 
 {} & -9.005 ($d$) & 0.114 & 0.909 & 0.865 &0.822 & 0.411 & 0.392\\
\hline
\end{tabular}
\label{T1}
\end{table*}

\begin{figure*}
\centerline{\includegraphics[width=17cm]{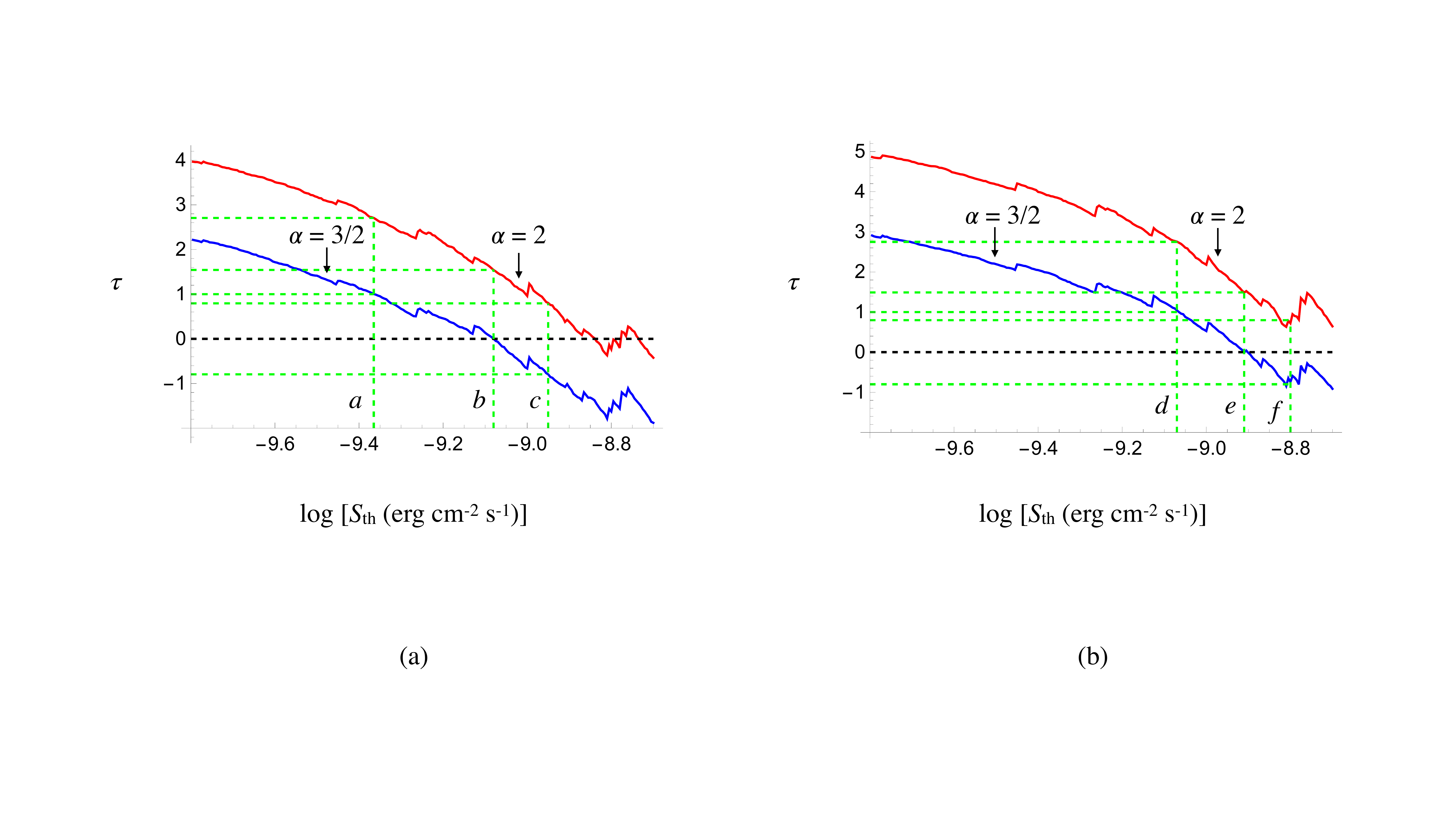}}
\caption{The Efron--Petrosian statistic $\tau$ versus the logarithm of the flux threshold $S_{\rm th}$ for the young pulsars.  The two curves are plotted for the values $2$ (red) and $3/2$ (blue) of $\alpha$.  (a) For the distances listed under YMW16 in the ANTF Catalogue.  The dashed vertical lines $a$, $b$ and $c$ (in green) designate the thresholds $\log S_{\rm th}=-9.365$, $\log S_{\rm th}=-9.08$ and $\log S_{\rm th}=-8.95$ which result in $\tau\vert_{\alpha=1.5}=1$, $\tau\vert_{\alpha=1.5}=0$ and $\vert\tau\vert_{\alpha=1.5}=\vert\tau\vert_{\alpha=2}$, respectively.  (b)  For the distances listed under NE2001 in the ANTF Catalogue.  The dashed vertical lines $d$, $e$ and $f$ (in green) designate the thresholds $\log S_{\rm th}=-9.066$, $\log S_{\rm th}=-8.911$ and $\log S_{\rm th}=-8.8$ which result in $\tau\vert_{\alpha=1.5}=1$, $\tau\vert_{\alpha=1.5}=0$ and $\vert\tau\vert_{\alpha=1.5}=\vert\tau\vert_{\alpha=2}$, respectively.}
\label{NLF6}
\end{figure*}

\begin{figure*}
\centerline{\includegraphics[width=17cm]{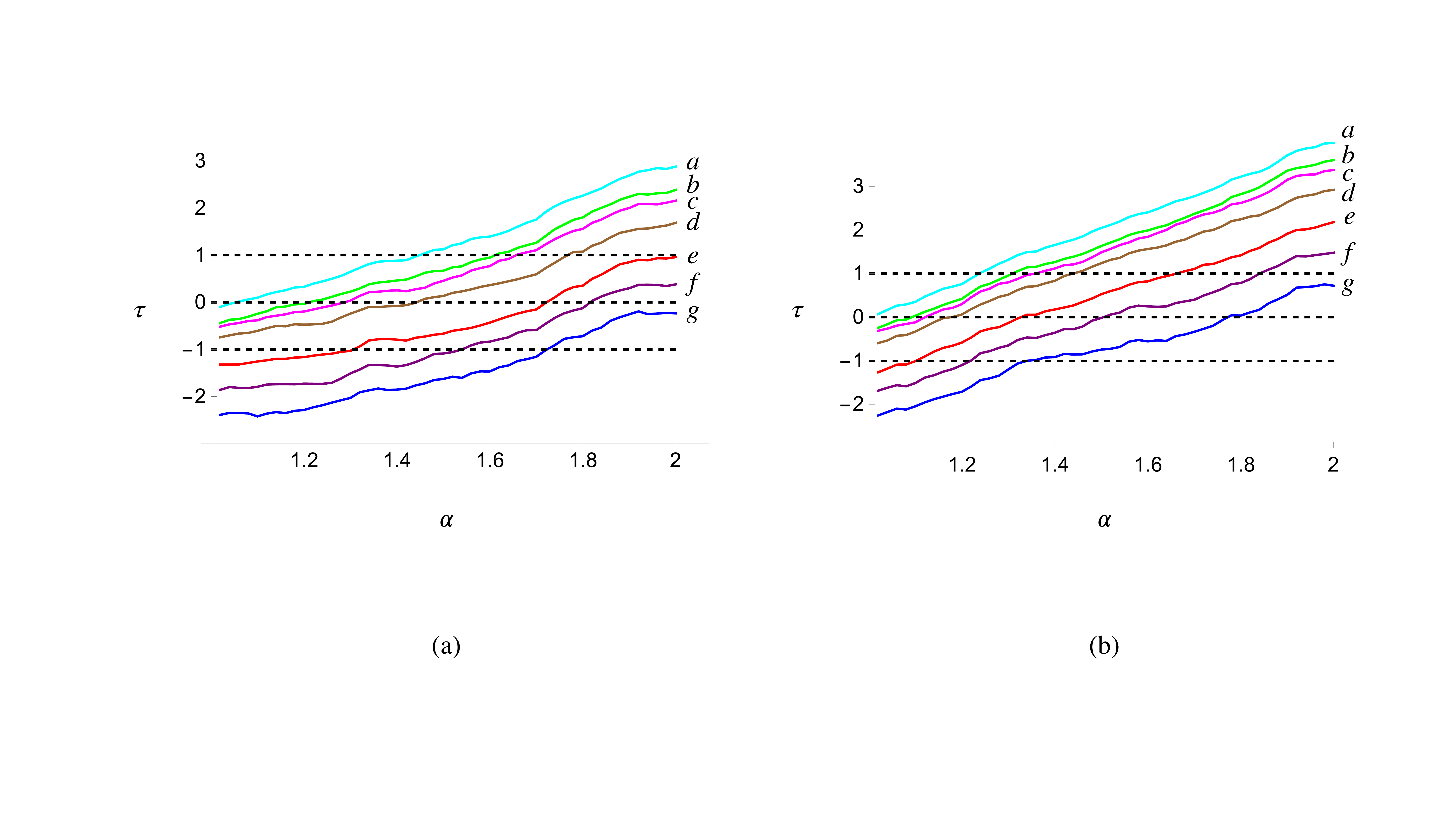}}
\caption{Contours of the function $\log S_{\rm th}(\alpha, \tau)$ over $-9.4\le\log S_{\rm th}\le-8.8$ at intervals of $0.1$ for the young pulsars.  (a)  Based on the distances listed under YMW16 in the ANTF Catalogue and for the following values of $\log S_{\rm th}$: -9.4 ($a$, cyan), -9.3 ($b$, green), -9.2 ($c$, magenta) , -9.1 ($d$, brown), -9 ($e$, red), -8.9 ($f$, purple), -8.8 ($g$, blue). (b) The same as in part (a) but based on the distances listed under NE2001 in the ATNF Catalogue.}
\label{NLF7}
\end{figure*}

\subsection{Test results for the young pulsars}
\label{subsec:young}

The counterparts of Figs~\ref{NLF4} and \ref{NLF5} for the data on young pulsars (Fig.~\ref{NLF2}) are plotted in Figs~\ref{NLF6} and \ref{NLF7}.  In Fig.~\ref{NLF6}, whose parts (a) and (b) are based on the YMW16 and NE2001 distances respectively, the flux thresholds designated by the letters $a$ to $f$ are chosen such that thresholds $a$ and $d$ each result in $\tau\vert_{\rm \alpha=1.5}= 1$, thresholds $b$ and $e$ each result in $\tau\vert_{\rm \alpha=1.5}= 0$ and thresholds $c$ and $f$ each result in $\vert\tau\vert_{\rm \alpha=1.5}= \vert\tau\vert_{\rm \alpha=2}$.  It can be seen from Figs~\ref{NLF6} and \ref{NLF7} and Table~\ref{T2} that in the case of young pulsars, too, there is a wide range of significance levels at which the hypothesis of independence of luminosity and distance can be rejected for $\alpha=2$ but not for $\alpha=3/2$.

\begin{figure*}
\centerline{\includegraphics[width=17cm]{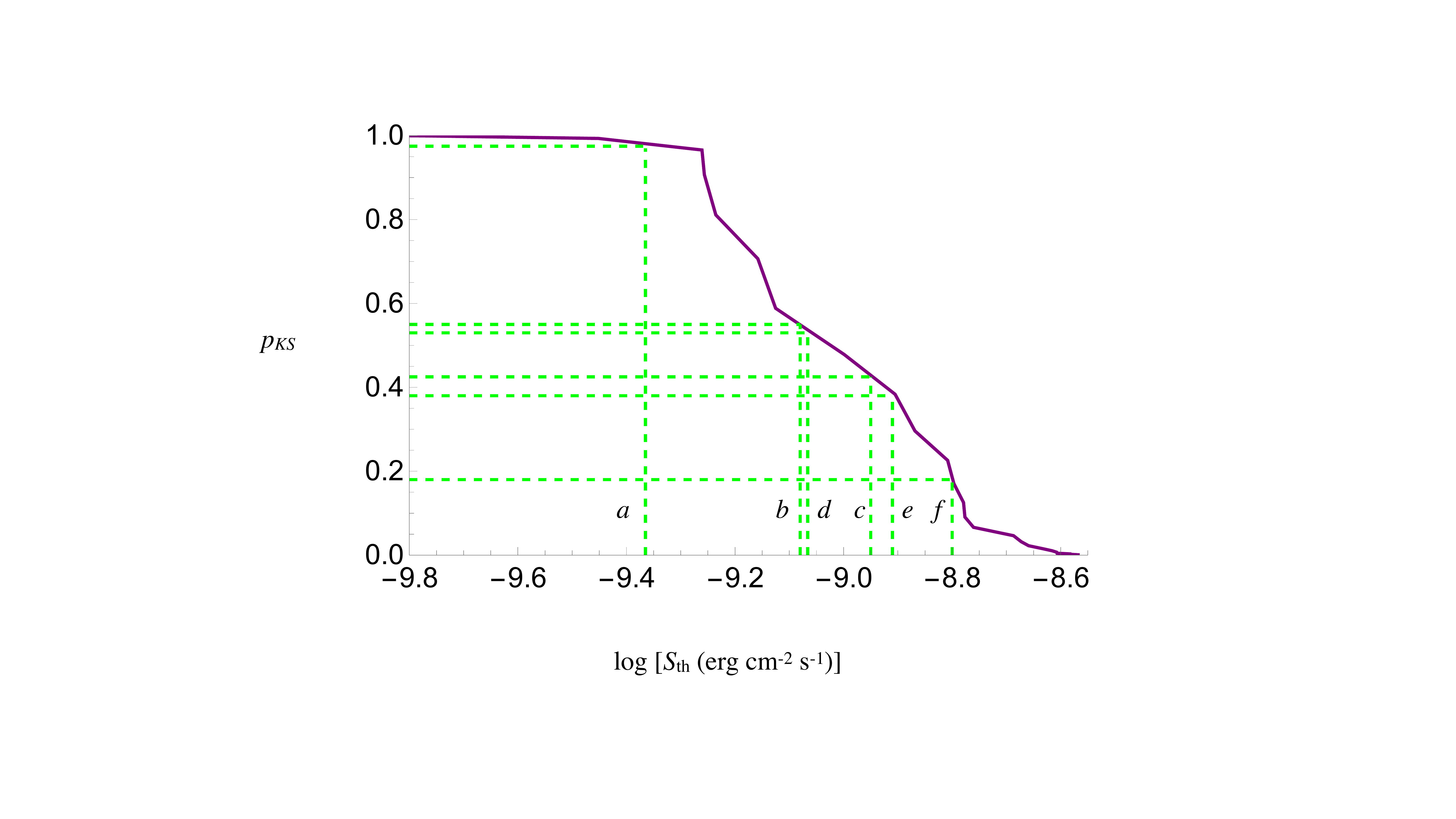}}
\caption{The Kolmogorov--Smirnov statistic $p_{KS}$ versus the logarithm of the flux threshold $S_{\rm th}$ for the young pulsars depicted in Fig.~\ref{NLF2}. The values of $p_{KS}$ corresponding to the flux thresholds $a$ ($\log S_{\rm th}=-9.365$), $b$ ($\log S_{\rm th}=-9.08$), $c$ ($\log S_{\rm th}=-8.95$), $d$ ($\log S_{\rm th}=-9.066$), $e$ ($\log S_{\rm th}=-8.911$) and $f$ ($\log S_{\rm th}=-8.8$), i.e.\ the thresholds designated by the same letters in Fig.~\ref{NLF6}, are respectively $0.975$, $0.55$, $0.425$, $0.53$, $0.38$ and $0.18$.}
\label{NLF8}
\end{figure*}

In this case, there do exist higher flux thresholds (higher than those designated $c$ in Fig.~\ref{NLF6}a and $f$ in Fig.~\ref{NLF6}b) which result in a greater value of $\vert\tau\vert$ for $\alpha=3/2$ than for $\alpha=2$.  However, the thresholds $c$ ($\log S_{\rm th}=-8.95$) and  $f$ ($\log S_{\rm th}=-8.8$) at which $\vert\tau\vert_{\alpha=1.5}=\vert\tau\vert_{\alpha=2}$ result in truncated data sets consisting, respectively, of only $82$ and $79$ elements.  The probability that these truncated data sets have the same origins as the $93$-element data sets depicted in Figs~\ref{NLF2}b and \ref{NLF2}c is comparatively low.  According to the Kolmogorov--Smirnov test, the probabilities $p_{KS}$ that the resulting $82$ and $79$-element data sets and the uncut $93$-element data set are drawn from the same distribution are $0.425$ and $0.18$, respectively (see Fig.~\ref{NLF8}).  In contrast, the probabilities that the truncated data sets resulting from the thresholds $a$ and $d$ in Fig.~\ref{NLF6} and the uncut $93$-element data set are drawn from the same distribution are given by $p_{KS}=0.975$ and $p_{KS}=0.53$, respectively (see Fig.~\ref{NLF8}).  It is essential that the observationally obtained data set and the part of it that lies above the chosen flux threshold could be regarded as drawn from the same distribution: from the unknown distribution that is complete over all values of the flux density. 

\begin{table*}
\centering
\caption{The Efron--Petrosian statistic $\tau$ and its associated $p$ and $p_{\rm perm}$ values for $\alpha=1.5$ and $2$ versus the flux threshold $S_{\rm th}$ for the young pulsars.  The rows YMW16 and NE2001 refer to the alternative distances listed in the ATNF Pulsar Catalogue and the letters $a$--$f$ refer to the flux thresholds shown in Fig.~\ref{NLF6}.  The last column shows the Kolmogorov--Smirnov statistic $p_{KS}$ associated with each flux threshold.}
\label{tab:landscape}
\begin{tabular}{lcccccccc}
\hline
${}$ & $\log [S_{\rm th}$ & $\tau\vert_{\alpha=1.5}$  & $p\vert_{\alpha=1.5}$ & $p_{\rm perm}\vert_{\alpha=1.5}$ & $\tau\vert_{\alpha=2}$ & $p\vert_{\alpha=2}$ & $p_{\rm perm}\vert_{\alpha=2}$ & $p_{KS}$\\
 {} & (erg cm$^{-2}$ s$^{-1})]$ & & & & & & &\\
\hline
 {YMW16} & -9.365 ($a$)  & 1 & 0.317 & 0.315 & 2.706 & 0.007 & 0.010 & 0.975\\
 {} & -9.080 ($b$) &  0 & 1 & 1 &  1.541    & 0.124 & 0.125 & 0.550\\
 {} & -8.950 ($c$) & -0.795 & 0.427 & 0.406 & 0.795 & 0.427 & 0.428 & 0.425\\
 \hline
 {NE2001} & -9.066 ($d$) & 1 & 0.317 & 0.309  & 2.75 & 0.006 & 0.001 & 0.53\\ 
 {} & -8.911 ($e$) & 0 & 1 & 1 &1.49 & 0.136 & 0.134 & 0.38\\
 {} & -8.800 ($f$) & -0.78 & 0.435 & 0.416 &0.78 & 0.435 & 0.455 & 0.18\\
\hline
\end{tabular}
\label{T2}
\end{table*}

\subsection{The effect of observational errors on the test results}
\label{subsec:errors}

Even if present, a purely systematic error in the estimates of distance and/or flux would not alter the results reported in Sections~\ref{subsec:millisecond} and~\ref{subsec:young}: the value of the Efron--Petrosian statistic does not change if we make monotonically increasing transformations on the values of distance and/or flux~\citep{EF1992}.  For instance, the dependence of $\tau$ on $\alpha$ (depicted by the curves in Fig.~\ref{NLF5}) remains exactly the same if the distances in the data sets shown in Figs~\ref{NLF1} and \ref{NLF2} are all multiplied by a positive factor.  In this section we perform a Monte Carlo simulation with $10^3$ random samplings to assess the effect on the test results of (the known) random errors in the estimates of flux. 

Fluxes are listed in the Fermi-LAT 12-year Catalog each with an uncertainty $\pm\sigma$~\citep{Abdollahi2022}.  The distribution of the error in the listed value of flux, $\mu$, can accordingly be modelled by a Gaussian probability density:
\begin{eqnarray}
f(x)&=&\sqrt{\frac{2}{\pi}}\sigma^{-1}\left[1+{\rm erf}\left(\frac{\mu}{\sqrt{2}\sigma}\right)\sigma\right]^{-1}\exp\left[-\frac{(x-\mu)^2}{2\sigma^2}\right],\nonumber\\
&&\qquad\qquad\qquad\qquad\qquad\qquad\qquad 0<x<\infty,
\label{E8}
\end{eqnarray}
where erf and H denote the error function and the Heaviside step function, respectively. 

\begin{figure*}
\centerline{\includegraphics[width=17cm]{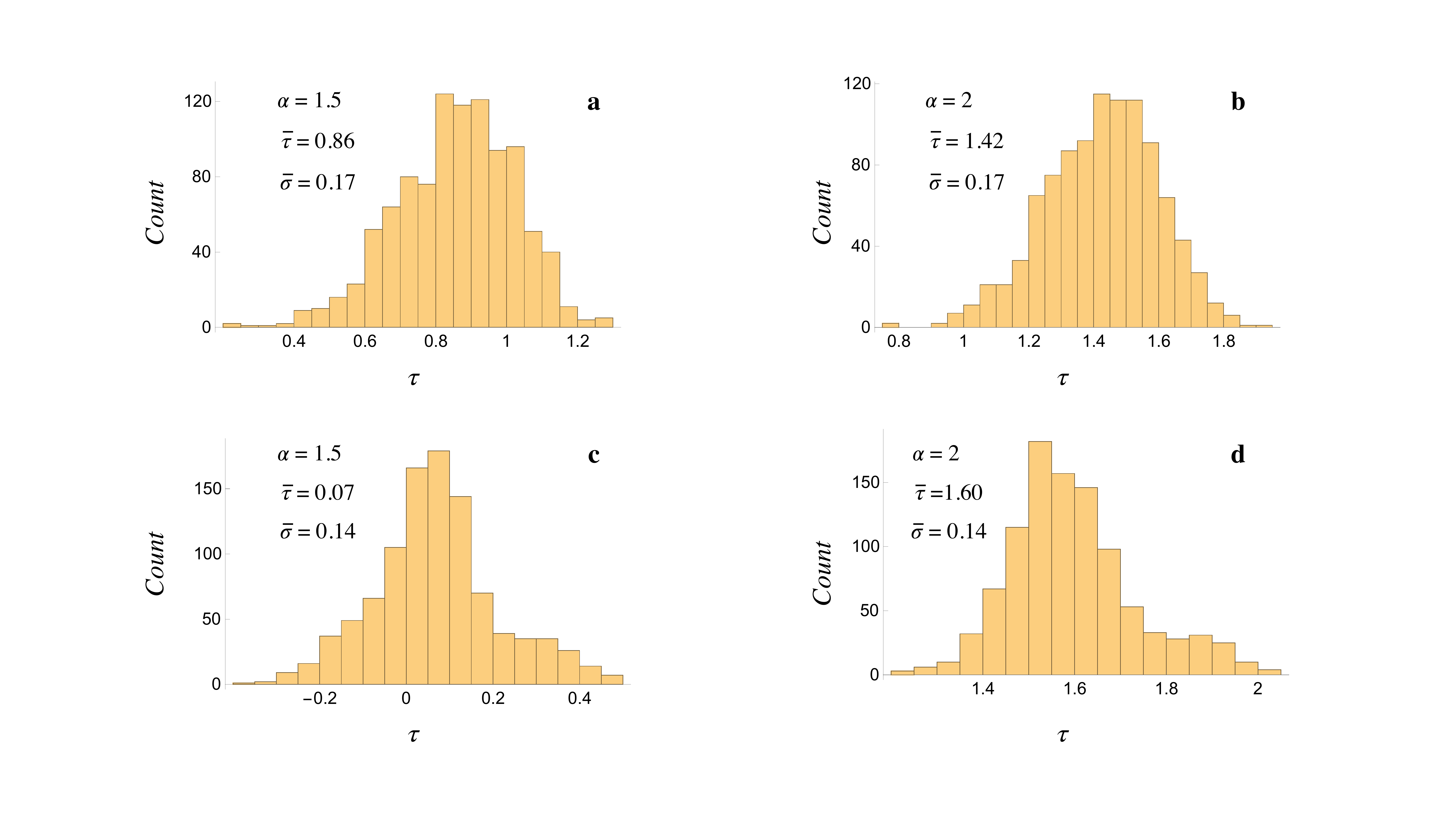}}
\caption{The histograms (obtained by the Monte Carlo simulation described in Section~\ref{subsec:errors}) of the distributions of the values of the Efron--Petrosian statistic $\tau$ that result from the inclusion of observational errors.  The distributions for the millisecond pulsars are plotted in parts ${\bf a}$ and ${\bf b}$ and those for the young pulsars in parts ${\bf c}$ and ${\bf d}$ of this figure.  The computations are based on the YMW16 distances and on the flux threshold $\log S_{\rm th}=-9.112$ (designated $a$ in Fig.~\ref{NLF4}a) for the millisecond pulsars and on the flux threshold $\log S_{\rm th}= -9.08$ (designated $b$ in Fig.~\ref{NLF6}a) for the young pulsars.  The mean and standard deviation of each distribution are denoted by $\bar\tau$ and $\bar\sigma$, respectively.}
\label{NLF9}
\end{figure*}

The sampling domain of the Monte Carlo method we use consists of the collection of the intervals $\mu-\sigma\le x  \le \mu+\sigma$ which enclose each flux coordinate (whose listed value we have denoted by $\mu$) of the elements of the data sets shown in Figs~\ref{NLF1} and~\ref{NLF2}.  In a given sampling, the flux coordinate of each element of these data sets is replaced by a randomly chosen value of $x$ from the probability distribution $f(x)$ over the interval $\mu-\sigma \le x \le \mu+\sigma$, where $\mu$ and $\sigma$ are the parameters of the flux coordinate of the data point in question as they appear in the Fermi-LAT 12-Year Catalog.  The modified data set thus obtained in a given sampling is then used, in conjunction with a choice of the flux threshold $S_{\rm th}$, to calculate the Efron--Petrosian statistic $\tau$ for both $\alpha=2$ and $\alpha=3/2$.  Repeating this procedure $10^3$ times and aggregating the two resulting sets of values of $\tau$ for differing values of $\alpha$, we arrive at the distributions shown in Fig.~\ref{NLF9} in the case of YMW16 distances.  These, as well as the corresponding distributions for NE2001 distances and other flux thresholds, show that the uncertainties in the estimates of pulsar fluxes, though bringing about a spreading of the values of $\tau$, do not alter the conclusions reached in Sections~\ref{subsec:millisecond} and \ref{subsec:young}.

In the case of the millisecond pulsars, the mean and standard deviation of the $\tau$-distribution for $\alpha=2$ in part ${\bf b}$ of Fig.~\ref{NLF9} have the values $1.42$ and $0.17$, respectively, so that the uncertainties in the estimates of flux result in replacing the value of the Efron--Petrosian statistic found in Section~\ref{subsec:millisecond}, i.e.\ $1.45$, by the interval $1.25 \le \tau \le 1.59$.  According to equation~(\ref{E7}), the $p$-values corresponding to this range of values of $\tau$ occupy the interval $0.11\le p \le0.21$.  In contrast,  the mean and standard deviation of the $\tau$-distribution for $\alpha=3/2$ in part ${\bf a}$ of Fig.~\ref{NLF9} have the values $0.86$ and $0.17$, respectively, so that the uncertainties in the estimates of flux result in replacing the value of the Efron--Petrosian statistic found in Section~\ref{subsec:millisecond}, i.e.\ $1$, by the interval $0.69 \le \tau \le 1.03$.  The $p$-values corresponding to this range of values of $\tau$ occupy the interval $0.30\le p \le0.49$. 

In the case of the young pulsars, the mean and standard deviation of the $\tau$-distribution for $\alpha=2$ in part ${\bf d}$ of Fig.~\ref{NLF9} have the values $1.60$ and $0.14$, respectively, so that the uncertainties in the estimates of flux result in replacing the value of the Efron--Petrosian statistic found in Section~\ref{subsec:young}, i.e.\ $1.54$, by the interval $1.46 \le \tau \le 1.74$. The $p$-values corresponding to this range of values of $\tau$ occupy the interval $0.08\le p \le0.14$.  In contrast, the mean and the standard deviation of the $\tau$-distribution for $\alpha=3/2$ in part ${\bf c}$ of Fig.~\ref{NLF9} have the values $0.07$ and $0.14$, respectively, so that the uncertainties in the estimates of flux result in replacing the value of the Efron--Petrosian statistic found in Section~\ref{subsec:young}, i.e.\ $0$, by the interval $-0.07 \le \tau \le 0.21$.  The $p$-values corresponding to this range of values of $\tau$, which includes $\tau=0$, occupy the interval $0.83\le p \le1$.

\begin{figure*}
\centerline{\includegraphics[width=17cm]{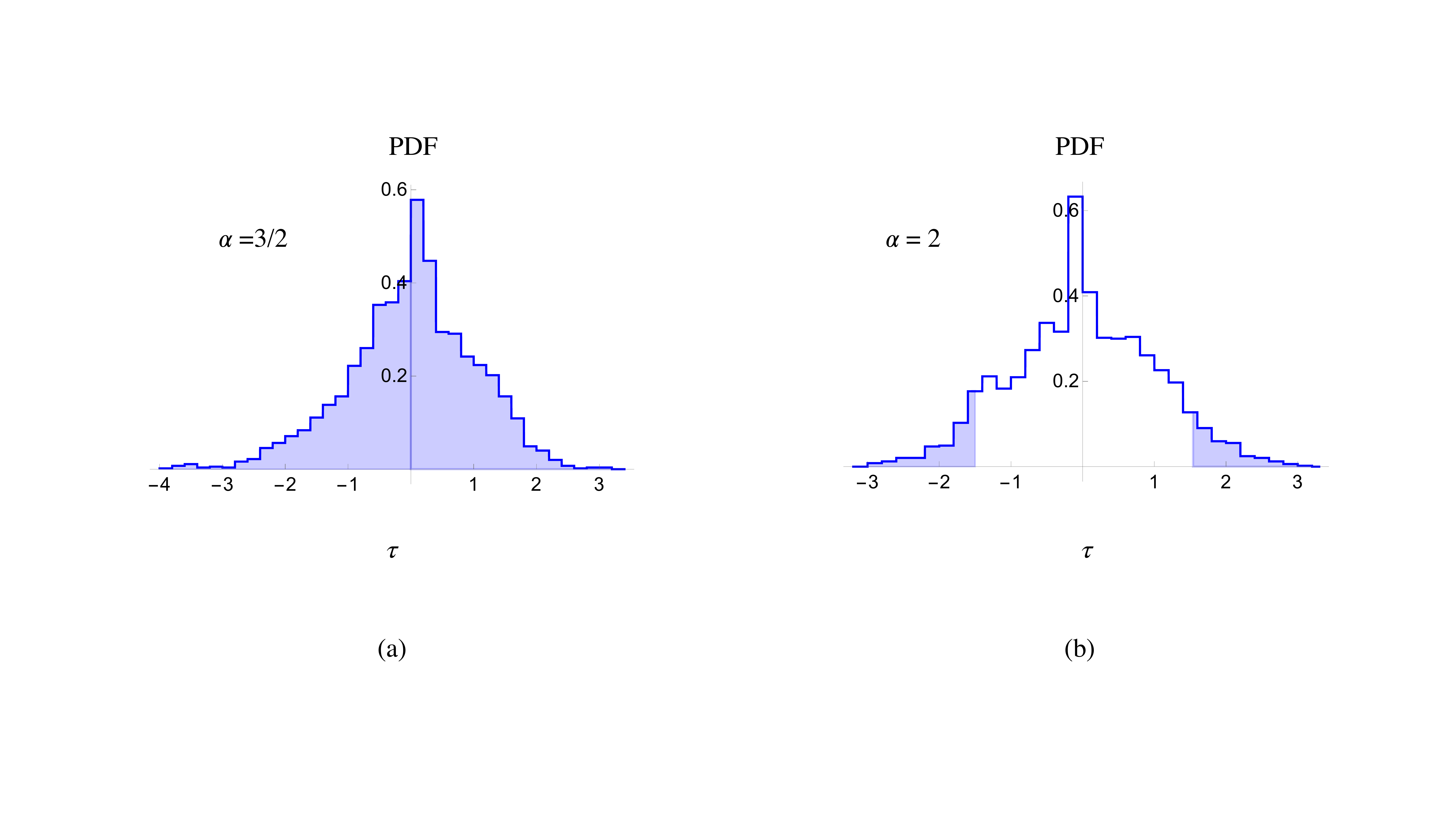}}
\caption{The normalized probability distribution functions of the Efron--Petrosian statistic $\tau$ for young pulsars (the data set shown in Fig.~\ref{NLF2}a) and the flux threshold $\log S_{\rm th}=-9.08$ (designated $b$ in Fig.~\ref{NLF6}a).  (a) The permutation distribution of $\tau$ for $\alpha=3/2$ based on an augmented data set consisting of $2749$ elements.  The shaded area under this distribution function yields $p_{\rm perm}=1$ for $\tau_{\rm obs}=0$, i.e.\ for the value of $\tau$ that is selected by the flux threshold $b$ and the blue curve in Fig.~\ref{NLF6}a.   (b) The permutation distribution of $\tau$ for $\alpha=2$ based on an augmented data set consisting of $2434$ elements.  The shaded area under this distribution function yields $p_{\rm perm}=0.125$ for $\tau_{\rm obs}=1.541$, i.e.\ for the value of $\tau$ that is selected by the flux threshold $b$ and the red curve in Fig.~\ref{NLF6}a.}
\label{NLF10}
\end{figure*}

\subsection{The effect of the finiteness of the size of the data set on the test results}
\label{subsec:size}

The only step in the analysis described in Section~\ref{subsec:statistic} at which the size of the data set is assumed to be large is where the resulting value of the Efron--Petrosian statistic $\tau$ is appraised by comparison with a normalized Gaussian distribution, i.e.\ the very last step at which the $p$ value associated with $\tau$ is calculated by means of equation~(\ref{E7}).  By relaxing that assumption, one can in fact apply the Efron--Petrosian method to data sets of any size: the normalized Gaussian distribution should be replaced, in the general case, by the normalized permutation distribution of the data set~\citep[see][Section 2.2]{EF1992}.

A permissible permutation of a truncated luminosity-distance data set, such as that shown in Fig.~\ref{NLF3}, is obtained by replacing an element $(\log D_i, \log L_i)$ of that data set by any one of the $N_i$ elements that constitute the set comparable to $(\log D_i, \log L_i)$.  The number of permuted data sets thus obtained from an original $n$-element data set is therefore given by $N=\sum_1^n N_i$.  In the case of the young pulsars shown in Fig.~\ref{NLF2}b, for example, the value of $N$ for the $83$-element data set corresponding to the threshold $\log S_{\rm th}=-9.08$ (designated $b$ in Fig.~\ref{NLF6}a) is $2749$ when $\alpha=3/2$ and $2434$ when $\alpha=2$.

If we calculate the Efron--Petrosian statistic $\tau$ for each one of the $N$ permuted data sets (corresponding to a given threshold and a given value of $\alpha$) and normalize the distribution of the resulting values of $\tau$ (by replacing each $\tau$ with $(\tau - {\bar\tau})/\sigma$, where ${\bar\tau}$ and $\sigma$ are the mean and standard deviation of the distribution), we arrive at a normalized probability distribution function representing the permutation distribution of $\tau$.  In the case of a data set with a limited number of elements, the observed value $\tau_{\rm obs}$ of $\tau$ (i.e.\ the value of $\tau$ determined by the original data set) should be appraised by comparison with this permutation distribution of $\tau$ instead of the normalized Gaussian distribution used at the end of Section~\ref{subsec:statistic}.  The size of the area that lies under a normalized permutation distribution in $-\infty<\tau\le-\vert\tau_{\rm obs}\vert$ and $\vert\tau_{\rm obs}\vert\le\tau<\infty$, here denoted by $p_{\rm perm}$, plays the role of the $p$-value defined in equation (\ref{E7})~\citep[see][Section 2.2]{EF1992}.

As an example, the normalized probability distribution functions of the Efron--Petrosian statistic for the $83$-element data set on young pulsars (Fig.~\ref{NLF2}a) that corresponds to the flux threshold $\log S_{\rm th}=-9.08$ (designated $b$ in Fig.~\ref{NLF6}a) and to the values $3/2$ and $2$ of $\alpha$ are shown in Fig.~\ref{NLF10}.  The mean and standard deviation of each of the shown distributions are given by $0$ and $1$, respectively.  The parameters $({\bar\tau},\sigma)$, i.e.\ the mean and standard deviation, of the un-normalized versions of these distributions have the values $(0,  0.150)$ for $\alpha=3/2$ and $(1.547, 0.141)$ for $\alpha=2$.   The shaded regions in this figure, whose areas determine $p_{\rm perm}\vert_{\alpha=3/2}$ and $p_{\rm perm}\vert_{\alpha=2}$, include all values of $\tau$ in $\tau\le-\vert\tau_{\rm obs}\vert$ and $\tau\ge\vert\tau_{\rm obs}\vert$ with $\tau_{\rm obs}=0$ for $\alpha=3/2$ and $\tau_{\rm obs}=1.541$ for $\alpha=2$ (see Fig.~\ref{NLF6}a).  In this case, $p\vert_{\alpha=3/2}=1$ and $p\vert_{\alpha=2}=0.124$ found in Section~\ref{subsec:young} on the basis of a normalized Gaussian distribution are respectively replaced by $p_{\rm perm}\vert_{\alpha=3/2}=1$ and $p_{\rm perm}\vert_{\alpha=2}=0.125$ (see Table~\ref{T2}).

Figure~\ref{NLF10} and its counterparts for other flux thresholds and alternative distances thus show that taking account of the finiteness of the sizes of the present data sets does not alter the outcomes of the test results obtained before: the values of $p_{\rm perm}$ only slightly differ from those of $p$ in most cases (see Tables~\ref{T1} and~\ref{T2}).

\section{Discussion}
\label{sec:conclusion}

The conclusion to be drawn from the above results is that the observational data in the Fermi-LAT 12-Year Catalog on gamma-ray pulsars are consistent with the dependence $S\propto D^{-3/2}$ of the flux densities $S$ of these pulsars on their distances $D$ at substantially higher levels of significance than they are with the dependence $S\propto D^{-2}$: a conclusion that agrees with those arrived at earlier~\citep{Ardavan2022b, Ardavan2022c} on the basis of the smaller data sets in the Second Fermi-LAT Catalog of Gamma-ray Pulsars~\citep{Abdo2013} and the McGill Magnetar Catalog~\citep{Olausen2014}.  To the list of observed features of the pulsar emission (brightness temperature, polarization, spectrum and profile with microstructure and with a phase lag between the radio and gamma-ray peaks) that the analysis of the radiation by the current sheet in the magnetosphere of a neutron star has decoded~\citep{Ardavan2021,Ardavan2022d}, we can therefore add another feature of this emission: the non-spherical decay of its high-frequency flux density.  

The violation of the inverse-square law encountered here is not incompatible with the requirements of the conservation of energy because the radiation process by which the superluminally moving current sheet in the magnetosphere of a neutron star generates the observed gamma-ray pulses is intrinsically transient.  Temporal rate of change of the energy density of the radiation generated by this process has a time-averaged value that is negative (instead of being zero as in a steady-state radiation) at points where the envelopes of the wave fronts emanating from the constituent volume elements of the current sheet are cusped~\citep[][Fig.~1 and Section 3.2]{Ardavan2021}.  The difference in the fluxes of power across any two spheres centred on the star is thus balanced by the change with time of the energy contained inside the shell bounded by those spheres~\citep[see][Appendix C, where this is demonstrated for each high-frequency Fourier component of a superluminally rotating source distribution]{Ardavan_JPP}.

Luminosities of gamma-ray pulsars are over-estimated when the decay of their flux density $S$ is assumed to obey the inverse-square law $S\propto D^{-2}$ instead of $S\propto D^{-3/2}$ by the factor $(D/\ell)^{1/2}$ (see equation~\ref{E1}).  The value of the scale factor $\ell$ is of  the same order of magnitude as the values of the light-cylinder radii of these pulsars~\citep[][Section 5.5]{Ardavan2021}.  Hence, the factor by which the luminosity of a $100$ ms gamma-ray pulsar at a distance of $2.5$ kpc is over-estimated is approximately $4\times10^6$.  Once this is multiplied by the ratio $\sim1/13$ of the latitudinal beam-widths of gamma-ray and radio pulsars (implied by the fraction of known pulsars that are detected in gamma-rays), we obtain a value of the order of $10^5$ for the over-estimation factor in question: a result that implies that the range of values of the correctly-estimated luminosities of gamma-ray pulsars is no different from that of the luminosities of radio pulsars.

For any given value of the angle between the magnetic and spin axes of the neutron star, there are four critical colatitudes~\citep[denoted by $\theta_{P1S}$, $\theta_{P2S}$, $\pi-\theta_{P1S}$ and $\pi-\theta_{P2S}$ in][]{Ardavan2021} with respect to the spin axis of the star along which the flux density of the radiation decays non-spherically.   The gradual change in the rate of decay of flux density with distance, from $D^{-3/2}$ to $D^{-2}$, away from a critical colatitude takes place over a latitudinal interval of the order of a radian.  But the latitudinal width of the tightly focused part of the radiation beam whose flux density decreases as $D^{-3/2}$ with distance $D$ has a much lower value: it is of the order of $(D/R_{lc})^{-1}$, where $R_{lc}$ is the radius of the star's light cylinder~\citep[][Section 5.5]{Ardavan2021}.  Hence, the solid angle centered on the neutron star within which the flux density of the present radiation decays at a significantly lower rate than that predicted by the inverse-square law is only a small fraction of $4\pi$.  

It follows that our lines of sight to only a small fraction of the total number of known neutron stars are expected to coincide with the privileged directions along which the flux density of the radiation we observe would decay as $D^{-3/2}$.  This is borne out by the fact that the known X-ray and gamma-ray emitting neutron stars comprise a small fraction of all observed pulsars.

\section*{Data availability}

The data used in this paper can be found at {\url{https://fermi.gsfc.nasa.gov/ssc/data/access/lat/12yr_catalog/}} and {\url{http://www.atnf.csiro.au/research/pulsar/psrcat}}.



\bibliographystyle{mnras}
\bibliography{Fermi.bib} 





\bsp	
\label{lastpage}
\end{document}